\documentclass[reprint,superscriptaddress,nofootinbib,amsmath,amssymb, aps]{revtex4-1}

\usepackage[a4paper, margin=2.2cm]{geometry}

\usepackage{bbm}
\usepackage{graphicx}
\usepackage{fancyhdr, color}
\usepackage{hyperref} 
\usepackage{float}
\usepackage{bm,bbm,color,mathrsfs,epstopdf}
\usepackage{hyperref,enumerate,enumitem, verbatim,amsmath,amsfonts}
\usepackage{graphicx,amssymb,amsthm,latexsym,revsymb,yfonts}
\usepackage{multirow,hhline,makecell,tabularx,braket}
\usepackage{dsfont,tcolorbox}
\usepackage[normalem]{ulem} 

\newtheorem{theorem}{Theorem}[section]
\newtheorem{definition}{Definition}[section]

\def\toto{\xrightarrow[\rm TO]~}

\newcommand{\hot}{\textup{Hot}}
\newcommand{\cold}{\textup{Cold}}

\newcommand{\battery}{\textup{W}}

\newcommand{\Wext}{W_\textup{ext}}

\newcommand{\noisy}{\textup{Noisy}}

\newcommand{\TO}{\textup{TO}}

\newcommand{\eig}{\textup{eig}}

\newcommand{\tr}{\mathop{\mathsf{tr}}\nolimits}

\newcommand{\stcold}{\textup{Cold}}
\newcommand{\me}{\mathrm{e}}

\newcommand{\proj}[1]{|#1\rangle\langle#1|}
\newcommand{\ketbra}[2]{|#1\rangle\langle#2|}

\newcommand{\be}{\begin{equation}}
\newcommand{\ee}{\end{equation}}

\theoremstyle{definition}

\newcommand{\ho}{{\rm Hot}}
\newcommand{\co}{{\rm Cold}}
\newcommand{\ma}{{\rm M}}
\newcommand{\bat}{{\rm W}}
\newcommand{\ext}{{\rm ext}}
\def\ba#1\ea{\begin{align}#1\end{align}} 

\usepackage{xparse}
\usepackage{times}
\newsavebox{\fminipagebox}
\NewDocumentEnvironment{fminipage}{m O{\fboxsep}}
{\par\kern#2\noindent\begin{lrbox}{\fminipagebox}
		\begin{minipage}{#1}\ignorespaces}
		{\end{minipage}\end{lrbox}%
	\makebox[#1]{%
		\kern\dimexpr-\fboxsep-\fboxrule\relax
		\fbox{\usebox{\fminipagebox}}%
		\kern\dimexpr-\fboxsep-\fboxrule\relax
	}\par\kern#2
}

\newcolumntype{M}[1]{>{\centering\arraybackslash}m{#1}}
\newcommand{\iden}{\mathbbm{1}}
\newcommand{\sgn}{\mathrm{sgn}}

\newcommand{\partthree}{{\color[rgb]{0.4,0.2,0.9} \sc Qbook:PartIII}}
\newcommand{\chintqt}{{\color[rgb]{0.4,0.2,0.9} \sc Qbook:Ch.0}}
 \newcommand{\chmcons}{{\color[rgb]{0.4,0.2,0.9} \sc Qbook:Ch.30}} 

\begin{document}

\fancyhead[R]{}

\title{Resource theory of quantum thermodynamics: Thermal operations and Second Laws}
\author{Nelly Huei Ying Ng}
\affiliation{Dahlem Center for Complex Quantum Systems, Freie Universit\"{a}t Berlin, 14195 Berlin, Germany}

\author{Mischa Prebin Woods}
\affiliation{Institute for Theoretical Physics, ETH Z\"{u}rich, Wolfgang-Pauli-Str. 27, 8093 Z\"{u}rich, Switzerland}

\date{\today}
\begin{abstract}
Resource theories are a generic approach used to manage any valuable resource, such as entanglement, purity, and asymmetry.
Such frameworks are characterized by two main elements: a set of predefined (free) operations and states,
that one assumes to be easily obtained at no cost. Given these ground rules, one can ask: what is achievable by using such free operations and states? This usually results in a set of state transition conditions, that tell us if a particular state $ \rho $ may evolve into another
state $ \rho' $ via the usage of free operations and states. We shall see in this chapter that thermal interactions can be modelled as a resource theory. The state transition conditions arising out of such a framework, are then referred to as ``second laws". We shall also see how such state transition conditions recover classical thermodynamics in the i.i.d. limit. Finally, we discuss how these laws are applied to study fundamental limitations to the performance of quantum heat engines.
\end{abstract}

\maketitle
\thispagestyle{fancy}

\section{Introduction} \label{sec:sectionlabel}
The term ``resource" refers to something that is useful, and most of the time, also scarce. This is because efforts are needed to create, store, and manage these things. What is useful, however, depends on the user: if one wants to send a message to a friend in a distant land, then digital communication channels are useful; if one would instead like to run a long computational code, then the computational power of his machine is the relevant resource.

Resource theories are conceptual, information-theoretic frameworks allowing one to quantify and manage resources. 
If an experimenter in his/her lab has a constrained ability to perform certain types of operations, then any initial state that cannot be created from such operations becomes valuable to him/her. A generic resource theory therefore is determined by two key elements: 
\begin{enumerate}[leftmargin=0.6cm,noitemsep]
	\item  a class of \emph{free operations} that are allowed to be implemented at no cost, 
	\item a class of \emph{free states} that one can generate and use at no cost\footnote{Any state which is not a free state is then called a resource state.}.
\end{enumerate}

Given the above operations and (an arbitrarily large number of copies of) free states that are assumed to be easily created, one can ask: if the experimenter possesses a quantum state $ \rho $, what are the set of states he/she can possibly reach by manipulation of $ \rho $, under the usage of these free operations and states? This produces the third aspect of resource theories, namely:
\begin{enumerate}[leftmargin=0.6cm,noitemsep]
	\setcounter{enumi}{2}
	\item \textit{state conversion conditions} that determine the possibility of inter-conversion between states, via the usage of free operations and free states. These conditions, either necessary or sufficient (sometimes both), are commonly phrased as \emph{monotones}, i.e. the transition is possible if a particular function $ f $ decreases in the transition $ \rho\rightarrow\rho' $, i.e. $ f(\rho)\geq f(\rho') $.
\end{enumerate}

A classic example of a resource theory comes from quantum information, in identifying \emph{entanglement} between quantum states as a resource.
Specifically, suppose that Alice and Bob are two distant parties who are capable of creating any local quantum states in their own labs and manipulating them via arbitrary local quantum operations. Furthermore, since classical communication is well established today (i.e. any transfer of non-quantum information such as an email encoded in a binary string ``001101...''), we may suppose that Alice and Bob can easily communicate with each other classically. 
Such additional classical communication allows them to create a joint, bipartite quantum state $ \rho_{\rm AB} $ which is correlated.
This set of operations is known as Local Operations and Classical Communication (LOCC) \cite{NieChu00Book,BBPS1996}, which defines a set of free operations and states. However, if Alice and Bob are allowed only such operations, it is then impossible for them to create any entanglement, so $ \rho_{AB} $ is separable. Therefore, any prior entangled state they share becomes a valuable resource, and it would be wise for them to manage it well, since it proves to be useful in various tasks. For example, if Alice and Bob share entanglement, they can use it together with classical communication to perform \emph{teleportation}~ | the transfer of a quantum state. A challenge is that many such tasks require entanglement in a pure, maximal form, namely in pure Bell states; while it is practically much easier to create states which are not maximally entangled. Therefore, the question of how can one \emph{distil} entanglement optimally is also frequently studied.
\begin{table*}
	\centering
	\begin{tabularx}{16.25cm}{ | M{2.6cm} | M{4.5cm} | M{3cm} | M{5.5cm} | }
		\hline
		\textbf{Resource Theory} & \textbf{Free operations} & \textbf{Free states} & \vspace{0.5cm}\textbf{State conversion conditions }\\[15pt]
		\hline 
		Entanglement (bipartite) \cite{NieChu00Book} & Local unitaries and classical communication & Any separable state $ \displaystyle\sum_i p_i \rho_{A,i}\otimes \rho_{B,i} $  & \vspace{0.5cm}
		$ \proj{\psi}_{AB} \rightarrow\proj{\phi}_{AB}  $ iff majorization holds: $\eig(\psi_A)\prec \eig(\phi_A)$. This condition implies that $ S(\psi_A) \geq S(\phi_A) $, i.e. the von Neumann entropy is a monotone, however the former condition is strictly stronger. \\[15pt] 
		\hline 
		Asymmerty w.r.t. a group G  \cite{marvian2013theory}& 
		Any quantum channel $ \mathcal{E} $ such that for any unitary representation $ U_G $, $ g\in G $, $ \mathcal{E}\left[U_g(\cdot)U_g^\dagger\right] = U_g \mathcal{E}(\cdot)U_g^\dagger$. & 
		Any state $ \rho $ such that $ \forall\, U_g $,  $ U_g\rho U_g^\dagger = \rho $ &  
		\vspace{0.5cm} For the symmetric group given by $ {\rm Sym}_G(\rho) = \left\lbrace g\in G: U_g \rho U_g^\dagger = \rho \right\rbrace $, then $ \rho\rightarrow\rho' $ only if $ {\rm Sym}_G(\rho) \subseteq {\rm Sym}_G(\rho') $.\\ [15pt]
		\hline 
		Coherence w.r.t. a basis \cite{baumgratz2014quantifying} $\lbrace |i\rangle \rbrace$ &  
		Incoherent operations, i.e. any quantum channel $ \mathcal{E} $ such that for any $ \rho\in\mathcal{I} $, then $ \mathcal{E}(\rho)\in\mathcal{I} $; where $ \mathcal{I} $ is the set of states diagonal w.r.t. basis $ \left\lbrace \ket{i}\right\rbrace $.&  
		Any $ \rho \in\mathcal{I}$ diagonal  in the basis $ \left\lbrace \ket{i}\right\rbrace $ &  
		\vspace{0.5cm} For the relative entropy of coherence given by $ \mathbf{C}_{RE} (\rho) = \displaystyle\min_{\sigma\in\mathcal{I}} D(\rho\|\sigma)$, then $ \rho\rightarrow\rho' $ only if 
		$ \mathbf{C}_{RE} (\rho) \geq \mathbf{C}_{RE} (\rho') $. \\[15pt]
		\hline 
		Purity & Unitary operations & Maximally mixed states & \vspace{0.3cm}Majorization: $ \rho\rightarrow\rho' $ iff $ \eig(\rho)\succ\eig(\rho') $ (Eq.~\eqref{eq:majorization}).  \\ [12pt]
		\hline 
		Thermodynamics & Energy-preserving unitary operations & Gibbs states (see Equation \eqref{eq:thermalstateGeneric}) & \vspace{0.3cm}Thermo-majorization (see Theorem \ref{thm:TMforTO}) \\ [8pt]
		\hline 
	\end{tabularx}
	\caption{A comparison of various resource theories. A quantum channel is a completely positive, trace preserving map.}\label{table:RTs}
\end{table*}
In Ref.~\cite{BBPS1996}, it was shown that if Alice and Bob have $n$ copies of a partially entangled state $ \rho_{\rm AB}^{\otimes n} $, one can, via LOCC, concentrate the amount of entanglement by producing $m$ copies of Bell states (where $m$ is smaller than $n$). Since its pioneering introduction, much experimental progress has been pursued, aiming to manage entanglement resources for the use of long-distance quantum communication \cite{kwiat2001experimental,takahashi2010entanglement,
	vollbrecht2011entanglement}.
In recent years, quantum resource theories have been studied not only in the generality of its mathematical framework \cite{brandao2015reversible,gour2008resource}, but also in particular those related to entanglement theory \cite{plbnio2007introduction,RevModPhys.81.865}, coherent operations \cite{baumgratz2014quantifying,winter2016operational}, or energetically in thermodynamics \cite{horodecki_reversible_2003,brandao2013resource}. In fact, the application of this framework in modelling thermal interactions has produced perhaps one of the most fundamental paradigm of quantum thermodynamics, inspiring the interpretation of many information-theoretic results to study thermodynamics for finite-sized quantum systems. 

Before we focus on the thermodynamic resource theory framework, let us consider once again the grand scheme of resource theories: Table \ref{table:RTs} summarizes and compares the main characteristics for various different quantum resource theories.
Entanglement theory is the most extensively studied case; however, due to great similarities in the mathematical framework, results can often be extended to the other resource theories as well. In the paradigm of entanglement resource theory: if one considers the set of LOCC operations as free operations, and separable states as free states, then any state that contains entanglement is a resource. In the next few sections, we shall identify the basic elements in a thermodynamic resource theory framework.

\section{Building the thermodynamic resource theory (TRT)}

As we have seen in the introduction \chintqt, thermodynamics is a theory concerning the change in energy and entropy of states in the presence of a heat bath. A special case in the theory, is when the heat bath has a fully degenerate Hamiltonian. In such cases, one cannot exchange energy with the heat bath, thus all operations consist in changes in entropy only. Before considering the full thermodynamic setting with arbitrary heat baths in subsection \ref{sub:TO}, it is illustrative to study this toy model first in subsection \ref{sub:NO}. We will then finalise this section with a discussion on the difficult topic of defining work in such small scale quantum systems in subsection \ref{Defining Work}, and the relation to other operations in subsection \ref{Relation to other operations}.

\subsection{Noisy Operations (NO)}\label{sub:NO}\vspace{-0.3cm}
\emph{Noisy operations}, which is perhaps the simplest known resource theory, is characterized by the following:
\begin{enumerate}[leftmargin=0.6cm,noitemsep]
	\item Free states are maximally mixed states of arbitrary (finite) dimension, with the form $ \frac{\iden}{d} $,
	\item All unitary transformations and the partial trace are free operations.
\end{enumerate}
Here the resource is \emph{purity}, which we shall quantify later. 
One can see intuitively why this is so: free states are those which have maximum disorder (entropy); on the other hand, unitary transformations preserve entropy, while the partial trace is an act of forgetting information, so these operations can never decrease entropy. The higher the purity of a quantum state, the more valuable it would be under noisy operations.

NOs are concerned only about the information carried in systems; instead of energy. For this reason, it has also been referred to as the \emph{resource theory of informational nonequilibrium}~\cite{gour2015resource}, or the \emph{resource theory of purity}. This toy model for thermodynamics was first described in \cite{horodecki_reversible_2003} and has its roots in the problem of exorcising Maxwell's demon~\cite{Bennett1982,landauer1961irreversibility},
building on the resource theory of entanglement manipulations \cite{BBPS1996,BBPSSW1996,thermo-ent2002,DevetakHW2005-resource,alicki2004thermodynamics}.

Formally, a transition $ \rho_S\xrightarrow[\rm NO]~\rho_S' $ is possible if and only if there exists an ancilla $R$ (with dimension $d_R$) such that 
\begin{equation}\label{eq:NO}
\rho_S' := \mathcal{E}_\noisy (\rho_S) := \tr \left[U_{SR}  \left(\rho_S\otimes\frac{\iden_R}{d_R}\right) U_{SR}^\dagger\right].
\end{equation}
Note that also since only unitaries $ U_{SR} $ are allowed, $\mathcal{E}_\noisy$ always preserves the maximally mixed state $\rho_S = \frac{\iden_S}{d_S}$, i.e. it is a unital channel.
Furthermore, since all unitaries can be performed for free, it is sufficient to consider only the case where $\rho_S,\rho_S'$ are diagonal in the same basis. Otherwise, one may simply define a similar NO, namely a unitary on $ S $ that changes $\rho_S$ to a state which commutes with $\rho_S'$. Now, let us denote the eigenvalues of $ \rho_S $ and $ \rho_S' $ as probability vectors $ p=\eig(\rho_S) $ and $ p'=\eig(\rho_S') $ respectively. 
It has been shown \cite{gour2015resource} that the following statements are equivalent:
\vspace{-0.05cm}
\begin{enumerate}[leftmargin=0.6cm,noitemsep]
	\item There exists $\mathcal{E}_\noisy$ so that Eq.~\eqref{eq:NO} holds~\footnote{Up to an arbitrarily good approximation to the final state $\rho_S'$ for fixed dimensional $\mathcal{H}_S$, in operator norm}.
	\item There exists a bistochastic matrix $ \Lambda_\noisy $ (namely, a matrix such that each row and each column add up to unity), such that $ \Lambda_\noisy p =p' $. 
\end{enumerate}
\vspace{-0.12cm}
This tells us that the state transition conditions for noisy operations are dependant solely in terms of the eigenvalues of initial and target quantum states.
An integral concept for understanding state inter-convertibility under NO is \textit{majorization}, which we now define.
\begin{definition}[Majorization]
	For two vectors $p$, $p'$, we say that $p$ majorizes $p'$ and write $p\succ p'$, if 
		\begin{equation}\label{eq:majorization}
	\sum_{i=1}^k p_i^\downarrow \geq \sum_{i=1}^k p_i'^\downarrow \quad \forall 1\leq k \leq d, 
	\end{equation}
	where  $p^\downarrow$ and $p'^\downarrow$ denote non-increasingly ordered permutations of $ p $ and $ p' $.
\end{definition}
The following theorem demonstrates the importance of majorization in resource theoretic thermodynamics:
\begin{theorem}[Birkhoff-von Neumann theorem, \cite{bhatia,hardy1952inequalities}]\label{thm:maj_bisto_Ttrans}
	For all $ n $-dimensional probability vectors $p$ and $ p'$, the following are equivalent:
	\begin{enumerate}[leftmargin=0.6cm,noitemsep]
		\item $p$ majorizes $p'$, namely  $p\succ p'$.
		\item $Ap = p'$ where $ A $ is a bistochastic matrix.
	\end{enumerate}
\end{theorem}
Quantum versions of the above theorem have been developed, and we refer the reader to a detailed set of notes on majorization in \cite{nielsen2002introduction}. The condition that $\rho_S$ can be transformed into $\rho_S'$ via noisy operations, is therefore a simple condition about the eigenvalues \cite{horodecki_reversible_2003}: 
\begin{equation}\label{eq:NOCond}
\rho_S\xrightarrow[\rm NO]~ \rho_S' \qquad\Longleftrightarrow \qquad p\succ p'.
\end{equation}
In fact, more can be said about the dimension of the ancilla R: it only needs to be as large as S, to enable the full set of possible transitions for any initial state $ \rho_S $ \cite{scharlau2018quantum}.

It is perhaps now interesting to note also that given the von Neumann entropy $$ S(\rho):=-\tr \rho\log\rho = H(\eig(\rho)),$$ where $ H(p) = -\sum_i p_i\log p_i $ is the Shannon entropy, Eq.~\eqref{eq:NOCond} also implies that $ S(\rho_S)\leq S(\rho_S') $. Such functions which invert the majorization order in Eq.~\eqref{eq:NOCond} are called Schur concave functions. 
This tells us that noisy operations always increase the entropy of system $ S $. However, majorization is also a much more stringent condition compared to the non-decreasing of entropy,
since there exists many other Schur concave functions.
Well-known examples are the $\alpha$-R{\'e}nyi entropies, which for $\alpha\in(-\infty,\infty)\backslash \{1\}$ are given by 
\begin{align}
	S_\alpha (\rho_S) &:= \frac{\sgn(\alpha)}{1-\alpha}\log \tr \rho_S^\alpha = H_\alpha(p),\label{eq:renyi S entropy}\\
	H_\alpha(p) &:= \frac{\sgn(\alpha)}{1-\alpha} \log \sum_i p_i^\alpha,\label{eq:renyi H entropy}
\end{align}
where the function $ \sgn (\alpha) $ gives 1 for $ \alpha \geq 0$ and -1 for $ \alpha <0 $. The von Neumann entropy and Shannon entropy can be viewed as special cases of Eqs. \eqref{eq:renyi S entropy},\eqref{eq:renyi H entropy}. Indeed, one can define $S_1 (\rho_S)$ and $H_1 (p)$ by demanding continuity in $\alpha$. Doing so, one finds that $S_1 (\rho_S)$ and $H_1 (p)$ give us the von Neumann and Shannon entropy respectively. All $\alpha$-R{\'e}nyi entropies are monotonically non-decreasing with respect to NOs. 

\subsection{Thermal Operations (TO)}\label{sub:TO}\vspace{-0.3cm}
In what sense are noisy operations similar to thermodynamical interactions? 

Note that the maximally mixed state (which we allow as free states in NOs) has a few unique properties: firstly, it is the state with the maximum amount of von Neumann entropy. It is also the state which is preserved by any noisy operation. This reminds one of the perhaps shortest way to describe the second law of classical thermodynamics: in an isolated system, disorder/entropy always increases. On the other hand, one can understand the emergence of the well-known canonical, Gibbs ensemble in thermodynamics as a statistical inference that assumes full ignorance (therefore maximum entropy) about the state, under the constraints of known macroscopic variables or conserved quantities such as average total energy. In particular, if the Hamiltonian of the system is fully degenerate, i.e. all microstates have the same amount of energy, then the Gibbs state is precisely the maximally mixed state.

On the other hand, if the system has a non-trivial distribution of energy levels in its Hamiltonian $ \hat H $, then under the constraint of fixed average energy, the state with maximum entropy is called the \emph{thermal/Gibbs state}, which has the form:
\begin{equation}\label{eq:thermalstateGeneric}
\tau^\beta = \frac{e^{-\beta \hat H}}{Z^\beta},
\end{equation}
where $\beta>0$ is the inverse temperature and is in a one-to-one correspondence with the mean energy of the Gibbs state, while $Z^\beta = \tr (e^{-\beta \hat H})$ is known as the \emph{partition function}
of the system. This quantity is directly related to crucial properties such as mean energy and entropy of a thermalized system, and here one sees that it is also the normalization factor for $ \tau^\beta $. 
Since the Gibbs state commutes with the Hamiltonian, it is also stationary under the evolution of the Hamiltonian.

Much work has been done to derive the emergence of Gibbs states in equilibration processes from the basic principles of quantum theory. The reader may refer to \partthree~
 for a more detailed explanation.
From this, we conclude that most quantum systems (i.e. generic Hamiltonians and initial states) will eventually equilibrate and tend towards the Gibbs state. 
This motivates the usage of Gibbs states as free states in the resource theory framework for thermodynamics. 		 
It has been noticed also that states of the form in Eq.~\eqref{eq:thermalstateGeneric} have a unique structure, called \emph{complete passitivity}~\cite{pusz_passive_1978}. This implies that even when provided with arbitrarily many identical copies, one cannot increase the mean energy contained in such states via unitary operations. The fact that energy preserving operations have to preserve the thermal state has been noted as early as \cite{Janzing2000}, while in \cite{2ndlaw}, by considering an explicit work-storage system while allowing only unitaries that commute with the global Hamiltonian, the thermal states defined in Eq.~\eqref{eq:thermalstateGeneric} enjoy a unique physical significance: they are the only states which cannot be used to extract work. This also implies that they are the only valid free states, since taking any other non-Gibbs state to be free states will allow arbitrary state transitions $\rho\rightarrow\sigma$, for any energy-incoherent $ \rho $ and $ \sigma $. This further justifies the usage of such Gibbs states as free states.

With these in mind, thermal operations were first considered in \cite{Janzing2000} and further developed in \cite{brandao2013resource,HO-limitations} in order to model the interaction of quantum systems with their larger immediate steady-state environment. The first restriction considered is that of \emph{energy-conserving unitary dynamics}: since systems are described by quantum states, the evolution should be described by unitary evolutions $U_{SR}$ across the closed system $S$ and bath $R$. Furthermore, the thermodynamical process described should preserve energy over the global system. Note that the demand that energy is conserved, implies that $U_{SR}$ commutes with the total Hamiltonian $\hat H_{SR} = \hat{H}_S\otimes\iden_R + \iden_S\otimes\hat{H}_R$. This still allows for energy exchange to occur between different systems, however, the interactions (governed by unitary dynamics) have to commute with the initial global Hamiltonian.

We are now ready to define thermal operations. Consider a system $S$ governed by Hamiltonian $\hat H_S$. For any $\beta\in\mathbb{R}_{\geq 0}$, a quantum channel $\mathcal{E}_\TO$ is a $\beta$-thermal operation if and only if there exists:
\begin{enumerate}[leftmargin=0.5cm,noitemsep]
	\item (free states) a Hamiltonian $\hat H_R$, with a corresponding Gibbs state
	\begin{equation}\label{eq:thermalstate}
	\tau_R^\beta = \frac{1}{Z_R^\beta} e^{-\beta\hat H_R}, ~Z_R^\beta = \tr \left(e^{-\beta\hat H_R}\right),
	\end{equation}
	\item (free operations) and a unitary $U$ such that $[U,\hat H_{SR}] = 0$, with
	\begin{equation}
	\mathcal{E}_\TO (\rho_S) = \tr_R \left[U_{SR} \left(\rho_S\otimes\tau_R^\beta\right) U_{SR}^\dagger\right].
	\end{equation}
\end{enumerate}
In the special case where $\hat H_S = \iden_S$ and $\hat H_R =\iden_R$, thermal operations $\mathcal{E}_\TO$ reduce to noisy operations.

A natural question to ask is what are the conditions on states $\rho_S,\rho'_S$ with Hamiltonian $\hat H_S$ such that 
they are related via a thermal operation, namely
$\rho_S\xrightarrow[\TO]~\rho'_S$? 
Ref.~\cite{brandao2013resource} first considered this question for asymptotic conversion rates, i.e. the optimal rate of conversion $R (\rho_S\rightarrow\rho_S') = \frac{m}{n}$ such that $\rho_S^{\otimes n}\rightarrow\rho_S'^{\otimes m}$, for the limit $n\rightarrow\infty$. Later, Ref.~\cite{HO-limitations} derived a set of majorization-like conditions which determine state transition conditions for a single copy of $\rho_S$ and $\rho_S'$. Such conditions are known as \emph{thermo-majorization}, and are shown to be necessary conditions for arbitrary state transitions via TO. Furthermore, when the target state $ \rho_S' $ commutes with $\hat H_S$, these conditions become also sufficient.

\begin{figure}[h!]
	\hspace{-1.5cm}\includegraphics[scale=0.57]{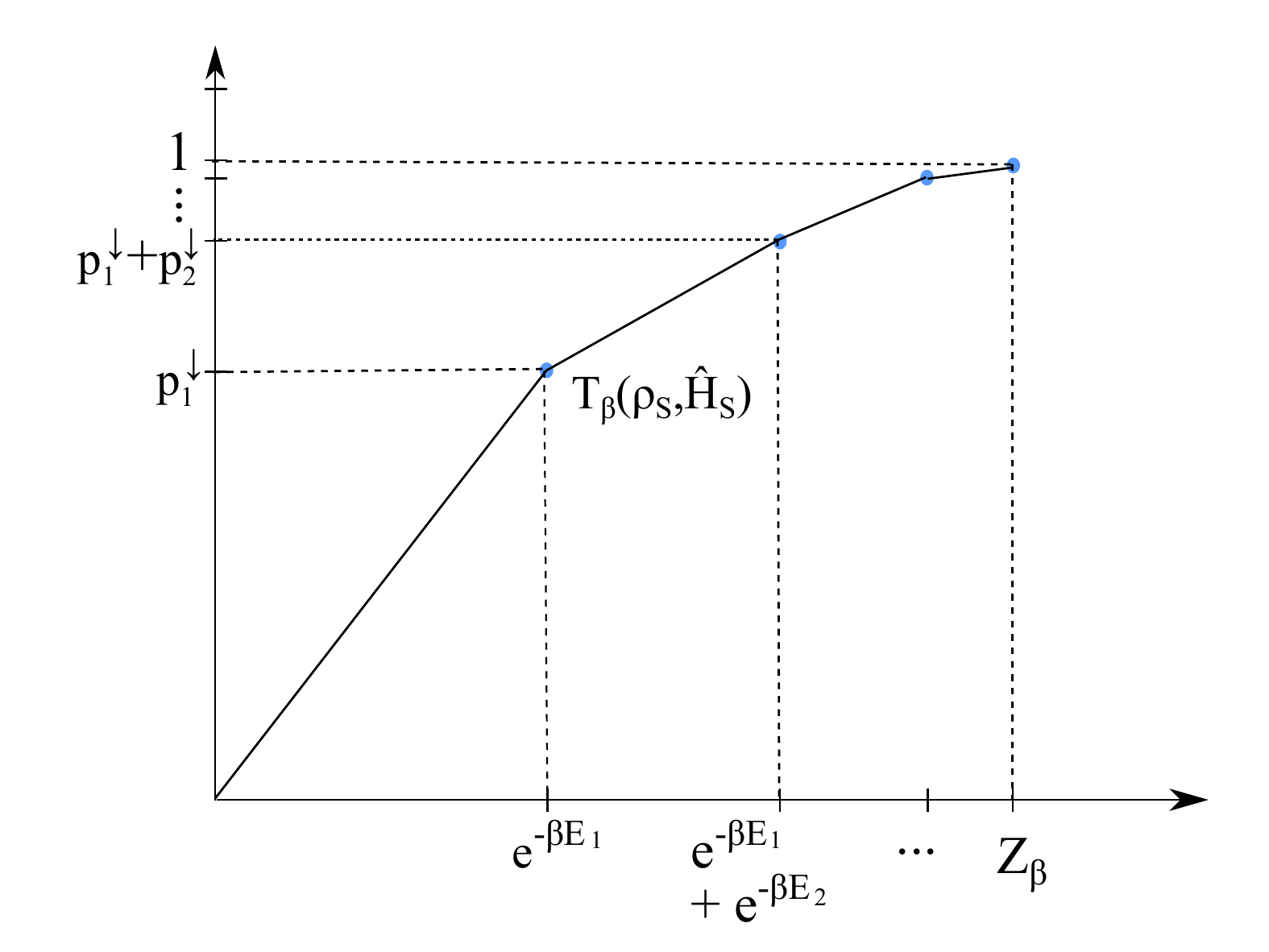}
	\vspace{-0.5cm}\caption{An example of a thermo-majorization curve of a state-Hamiltonian pair $T_\beta(\rho_S,\hat H_S)$.}\label{fig:tm}
\end{figure}

To describe these conditions, first let us explain a thermo-majorization curve. Consider any state-Hamiltonian pair $(\rho_S,\hat H_S)$ where $\rho_S$ has rank $d$ and commutes with $\hat H_S$. These states are commonly analyzed in the framework of thermodynamic resource theories, and are referred to as \emph{energy incoherent states}, or simply states that are \emph{block-diagonal}. Because $\rho_S$ and $\hat H_S$ commute, $ \rho_S $ is diagonalisable in a particular energy eigenbasis of $\hat H_S$, and can be written in the form
\begin{equation}\label{eq:E vals rhoS}
\rho_S = \sum_{E,g_E} p_{E,g_E} \proj{E,g_E}_S,
\end{equation}
where $p_{E,g_E}$ are eigenvalues of $\rho_S$ and $\{\ket{E,g_E}\}_{g_E}$, $E$ are the corresponding energy eigenvectors and values of $\hat H_S$ with degeneracy $g_E$. A thermo-majorization curve is then defined in Def.~\ref{def:TMC} (Box 1), and
\begin{figure*}
	\begin{fminipage}{\textwidth}\vspace{0.1cm}
		\begin{definition}[Thermo-majorization curve]\label{def:TMC}
			A $\beta$ thermo-majorization curve $T_\beta(\rho_S,\hat H_S)$ is defined by first ordering the eigenvalues of $\rho_S$ in Eq. \eqref{eq:E vals rhoS} to have $p^{\downarrow\beta} = (p_1^\downarrow,\cdots,p_d^\downarrow)$ with the corresponding energy eigenvalues $E_1, \cdots,E_d$ such that 
			\begin{equation}\label{eq:betaorder}
			p_1^\downarrow e^{\beta E_1} \geq p_2^\downarrow e^{\beta E_2} \geq \cdots p_d^\downarrow e^{\beta E_d}.
			\end{equation}
			Such an ordering is called $\beta$-ordering, which may be non-unique when parts of Eq.~\eqref{eq:betaorder} are satisfied with inequality. However, once the eigenvalues are ordered this way, a concave, piecewise linear curve called the \emph{thermo-majorization curve of $(\rho_S,\hat H_S)$} is uniquely defined: by joining all the points
			\begin{equation}\label{eq:TMD}
			\left\lbrace (0,0),~\left(e^{-\beta E_1},p_1^\downarrow\right), ~\left(e^{-\beta E_1}+e^{-\beta E_2},p_1^\downarrow+p_2^\downarrow\right),~\cdots, ~\left(\sum_{i=1}^d e^{-\beta E_i},\sum_{i=1}^d p_i^\downarrow\right) \right\rbrace.
			\end{equation}
		\end{definition}
		\vspace{0.03cm}
	\end{fminipage}
\end{figure*}
Fig.~\ref{fig:tm} shows an example of a thermo-majorization diagram defined by coordinates in Eq.~\eqref{eq:TMD}. Notice that if $ E_1=E_2=\cdots=E_d $, then the intervals on the $ x $-axis are equally spaced, and $ \beta $-ordering would be the ordering of eigenvalues $ p_i $ in a non-increasing manner. On the other hand, for such Hamiltonians, thermal operations reduce to noisy operations, and the state transition conditions to majorization. 
It is proven in \cite{HO-limitations}, that for states commuting with their Hamiltonian, the comparison of two thermo-majorization curves also dictate the possibility of state transition via thermal operations, as detailed by the following theorem:

\begin{theorem}\label{thm:TMforTO}
	Consider state-Hamiltonian pairs $(\rho_S,\hat H_S)$ and $(\rho_S',\hat H_S')$, such that $ [\rho_S,\hat H_S] = [\rho_S',\hat{H}_S']=0$. The state transition $\rho_S\xrightarrow[TO]~ \rho_S'$ can happen if and only if $T_\beta(\rho_S,\hat H_S)\geq T_\beta(\rho_S',\hat H_S')$, i.e. the $ \beta $ thermo-majorization curve of $(\rho_S,\hat H_S)$ lies above that of $(\rho_S',\hat H_S')$. In this case, we say that $ \rho_S $ thermo-majorizes $ \rho_S' $.
\end{theorem}

Note that for the state $ \tau_S^\beta = (Z_S^\beta)^{-1}\cdot  e^{-\beta\hat H_S}$, its thermo-majorization curve $ T_\beta (\tau_S^\beta,\hat{H}_S) $ simply forms a straight line with endpoints $ (0,0) $ and $ (Z_\beta, 1) $. Therefore, for any other block-diagonal state $ \rho_S $, Theorem \ref{thm:TMforTO} implies that $ \rho_S\toto\tau_S^\beta $ is always possible via a $ \beta $-thermal operation, in accordance with the laws of thermodynamics and our notion of a free state.

As a closing remark, the astute reader may have noticed that we have not commented on the conditions for state conversion under TO when the initial and (or) final states are energy-coherent. For arbitrary states $ \tilde\rho_S $, the corresponding thermo-majorization is defined to coincide with the dephased version of $ \tilde{\rho_S} $ in the energy eigenbasis; and this still provides necessary conditions for state transition \cite{lostaglio2015quantum}. So in conclusion, in general, for TOs it remains an open question as to what are the necessary and sufficient conditions characterizing transitions between arbitrary, energy-coherent states.

\subsection{Relation to other operations}\label{Relation to other operations}\vspace{-0.3cm}
\subsubsection{Gibbs preserving Maps (GPs)}
\vspace{-0.4cm}
Within the resource theory approach to quantum thermodynamics, and arguably further afield, the most generic model for thermal interactions are \emph{Gibbs preserving maps}. Understood literally, this means that the set of free operations is simply the set of all quantum channels that preserve the Gibbs state of some inverse temperature $ \beta $, i.e.
\begin{equation}
\mathcal{E}_{GP} \left(\tau_S^\beta\right) = \tau_S^\beta.
\end{equation}
One can view GPs as highlighting the ``bottomline'' of any model for thermodynamical interactions. For initial and final states which are block-diagonal, the set of allowed transitions via GPs coincide with thermal operations. However, for general quantum states, GPs may act on energy-incoherent initial states to create energy-coherent final states. This is not possible via thermal operations.

GPs are one of the less studied thermodynamic resource theory models, since there is no known explicit physical process that describes the full set of GPs. Nevertheless, the existence of a GP map can be phrased as a semi-definite problem, and this gives rise to straightforward necessary and sufficient conditions for such a map. In \cite{faist2017fundamental}, it has been shown that this condition can be phrased in terms of a new entropic quantity, that acts as a parent quantity for several entropic generalizations.

\subsubsection{Multiple conserved quantities}
\vspace{-0.4cm}
A quantum system may in general obey several conservation laws other than total energy, where the conserved quantities (such as spin, momentum etc) are represented by operators on the system which do not generally commute. In this case, one can also further restrict the set of free operations to conserve other constants of motion. Systems in this case tend to equilibrate to the generalized Gibbs ensemble instead of thermalizing to Eq.~\eqref{eq:thermalstateGeneric}. In \cite{halpern2015microcanonical,sandu_noncommute,lostaglio2017thermodynamic,Popescu2018}, such scenarios have been studied in order to model not only energetic/information exchanges in thermodynamics, but also including exchanges of other non-commuting observables. For further discussion, refer to chapter \chmcons. In another vein, necessary and sufficient conditions for arbitrary state transitions have been derived in \cite{Gour2017} for a set of operations called \emph{generalized thermal processes}, which also take into account such non-commuting variables. While it is unclear if such operations reduce to the set of TOs in the case where only energy is conserved, nevertheless, when the considered states are energy-incoherent, the necessary and sufficient conditions in \cite{Gour2017} do reduce to thermo-majorization.

\subsection{Defining Work}\label{Defining Work}\vspace{-0.3cm}
A central focus of thermodynamics is the consumption/extraction of work, which is the output/input of ordered energy to a system. Therefore, we must ask the question: in the context of TRTs, what does it mean to extract work? 

To gain some intuition for a rigorous formulation, let us recall how this is approached in classical thermodynamics. Work is often pragmatically pictured as the effect of storing potential energy on a system, for example designing a protocol involving a hanging weight, such that in the end, the weight undergoes a height difference $ \Delta x $. 
It is also a long-standing observation that one cannot extract work solely from a thermal reservoir; however given two reservoirs at distinct temperatures, one can design protocols to extract work. 
A common approach is to consider a heat engine, where a machine interacts with two different heat baths successively, and undergoes a cyclic process \cite{reif,adkins,schroeder,huang}. A physical example of such a ``machine'' could be a cylinder of ideal gas, where the volume can be changed with a piston. By coupling the piston to the weight, and allowing the machine to go through a series of isothermal/adabatic processes while interacting with two reservoirs, one may analyze the net energy flow in and out of this machine system, and calculate the energy output on the weight, while assuming that energy lost/dissipated (for example, via friction) is negligible.

In the quantum regime, earlier approaches \cite{egloffdahlsten,del2011thermodynamic,aaberg2013truly,Crooks1998,infoerasure} have considered different sets of operations. Some popular approaches include allowing for a mixture of 1) level transformations, which is the freedom to tune energy levels in the Hamiltonian and 2) thermalization. These operations are non-energy preserving in general.
Therefore, with each operation, one may refer to an amount of work done on/by the system \cite{Esposito2011}. This amount of work would be largely influenced by energy fluctuations in the system.
In particular, much discussion has gone into how one should differentiate work from heat \cite{hossein2015work,aaberg2013truly,definingworkEisert,gemmer2015single}. Although both contribute to a change in energy of the system, work stored is of an \emph{ordered} form, and therefore can be extracted and used, while heat is irreversibly dissipated/lost. Though a seemingly simple problem, there is no consensus among the community as to how work should be defined. We describe the main two different strategies in this section.

\subsubsection{Average work}\label{average work}\vspace{-0.3cm}
A popular way to quantify work is to model the weight as a quantum system (referred to as a battery) that undergoes a change $ \rho_W\rightarrow\rho_W' $ during a thermodynamic process \cite{extworkcorr,skrzypczyk2014work,PhysRevLett.88.050602}. Work is then defined as 
$ \Delta W = \tr(\hat H_W \rho_W') - \tr(\hat H_W \rho_W) $.
Since $ \Delta W $ is the average energy change, which is subjected to random processes (ex: thermalization), work is treated as a random variable. 
Studies have shown that for example, an optimal amount of average work (equal to the free energy of the system) can be drawn, or that Carnot efficiency can be achieved \cite{skrzypczyk2014work}.

The simplicity of this quantity makes it one of the first choices in quantifying work. However, when such a measure is used, it is crucial that the amount of entropy has to be separately analyzed, in order to show that heat contributions to the average energy increase are either negligible, or at least accounted for. This is because thermodynamic processes could, and often do produce work with fluctuations of the same order \cite{aaberg2013truly,Richens2016}. Such fluctuations are undesirable for two reasons: (1) they affect the reusability of the battery, i.e. one might not be able to extract the full amount of $ \Delta W $ out again, (2) the battery could have potentially been used as an ``entropy sink", 
where the average work cannot capture this effect, thus leading to an apparent violation of the second law.

To solve the latter problem, a recent approach is to further restrict the allowed operations, so that they satisfy \emph{translational invariance} on the battery \cite{aberg2016fully,alhambra2016second}. By this restriction, one can show that defining work as above still allows one to formulate refined versions of the second law. While this approach is conceptually a large step forward in justifying the reasonableness in using average energy as a work quantifier, several improvements await.
One of the main caveats is that the current battery models in such approaches have no ground state; which is a reasonable approximation only when the initial and final states of the battery have extremely high energy. Understanding the corrections that enter this picture when using physical battery models is therefore of high importance.
\subsubsection{Single-shot and deterministic work}\label{subsub:detwork}\vspace{-0.3cm}
Another common approach in TRTs \cite{HO-limitations, 2ndlaw} (although not restricted to resource theories), is to phrase the work extraction problem as a state transition problem on the battery. More concretely, one fixes a particular desirable, perhaps more physically motivated battery Hamiltonian $ H_W $, along with specific battery states $ \rho_W,\rho_W' $, and then consider the possibility of the state transition
\begin{equation}\label{eq:ssw}
\rho_S \otimes\rho_W \xrightarrow[X]~\rho_S'\otimes\rho_W',
\end{equation} 
for any arbitrary final state $ \rho_S' $ (one can consider $X$ for TOs, CTOs, or other processes as well).
Commonly used models of battery Hamiltonians include a two-level qubit with a tunable energy gap \cite{HO-limitations, 2ndlaw}, a harmonic oscillator \cite{skrzypczyk2014work}, or a system with quasi-continuous energy levels \cite{Woods2015}. For an explicit example, if we use the two-level qubit battery such that $ \hat H_W = \Wext \proj{1}_W $, then a transition from the state $ \rho_W = \proj{0}_W $ to $ \rho_W' = \proj{1}_W $ corresponds to extracting an amount of work equal to $ \Wext $. At the same time, we have required that we retain perfect knowledge of the final state of $ \rho_W' $, and no entropy increase occurs.

Most often, the explicit battery model (i.e. its Hamiltonian $ \hat H_W $) does not affect the amount of work stored/used. However, it does depend on the initial and final battery states $ \rho_W,\rho_W' $.

\section{Second laws of quantum thermodynamics}
The framework of thermal operations, presented so far, incorporates the zeroth and first law of thermodynamics very naturally. The zeroth law is established by noting that Gibbs states are singled out as special, free states that are uniquely characterized by a parameter $ \beta $. The first law comes in due to the requirement that only energy preserving unitaries are allowed. But where is our second law? Thermo-majorization governs state transitions, however it is very different from what we usually know as the second law, namely: when a system is brought into contact with a heat bath, its free energy 
	\begin{equation}\label{eq:free energy}
	F(\rho,\hat H):= \tr(\rho \hat H)-\beta^{-1}S(\rho),
	\end{equation}
	never increases. While we know that if $ \rho $ thermo-majorizes $ \sigma $, then $ F(\rho)\geq F(\sigma) $, thermo-majorization is however much stricter than just requiring free energy to decrease. Is there hope to bridge these two statements more closely? The answer is affirmative, by two steps: (1) allowing a catalyst, and (2) looking at approximate transitions in the i.i.d. limit.
\subsection{Catalytic TOs (CTOs)}\label{sub:CTO}\vspace{-0.3cm}
CTOs are extensions of TOs, where in addition to the same free resources and operations, catalysts that remain uncorrelated and unchanged are allowed. In other words, $ \rho_S\xrightarrow[\rm CTO]~\rho_S' $ is possible via a $ \beta $-CTO iff there exists 
\begin{enumerate}[leftmargin=0.4cm,noitemsep]
	\item (free states) a Hamiltonian $\hat H_R$, with a corresponding Gibbs state $ \tau_R^\beta $ as in Eq.~\eqref{eq:thermalstate},
	\item (catalysts) any additional finite-dimensional quantum state $ \omega_C $ with Hamiltonian $ \hat H_C $,
	\item (free operations) a global unitary $U_{SRC}$ such that $[U_{SRC},\hat H_{SRC}] = 0$, and
	\begin{equation}
	~~~\tr_R \left[U_{SRC} \left(\rho_S\otimes\tau_R^\beta\otimes\omega_C\right) U_{SRC}^\dagger\right] = \rho_S'\otimes\omega_C.
	\end{equation}
\end{enumerate}

There are transitions that cannot occur via TOs, but are made possible by a catalyst. An example is seen in classical thermodynamics: a system going through a heat engine cycle, interacting successively with multiple heat baths and outputting some work before returning to its original state. Mathematically, non-trivial examples can already be demonstrated for systems of dimension 4, with a 2-dimensional catalyst \cite{DP99}.
Similarly to the case of TO, for CTO we would like to have conditions on the states $\rho_S,\rho_S'$ and their Hamiltonian $\hat H_S$ for $ \rho_S\xrightarrow[\rm CTO]~\rho_S' $ to occur. It turns out that such conditions exist, and they show a deep connection with entropic measures used often in quantum information theory.

\subsection{Necessary conditions for arbitrary state transitions}\vspace{-0.3cm}
As earlier mentioned, necessary and sufficient conditions for state transitions under TO remains a large open question for TRTs. This is also the case for CTO. However, we know that since TOs are quantum channels that preserve the thermal state, one can derive easily necessary conditions for CTOs. In particular, if we have any function $ f(\rho||\sigma) $ where $ \rho,\sigma $ are density matrices, and if we know that $ f $ satisfies the data processing inequality, meaning that for any quantum channel $ \mathcal{E} $, and any $ \rho,\sigma $, we have
\begin{equation}\label{eq:DPI}
f(\rho||\sigma) \geq f(\mathcal{E}(\rho)||\mathcal{E}(\sigma)),
\end{equation}
then we can derive that $ f(\rho_S||\tau_S^\beta) $ is also monotonically decreasing under TOs, where $ \tau_S^\beta $ is simply the thermal state of the system, w.r.t. Hamiltonian $ \hat H_S $. Examples of such functions are the quantum R{\'e}nyi divergences:
\begin{equation}\label{eq:19}
\tilde D_\alpha(\rho||\sigma) =\begin{cases}
\frac{1}{\alpha-1} \log\tr(\rho^\alpha\sigma^{1-\alpha})&\alpha\in [0,1)\\
\log\left((\sigma^{\frac{1-\alpha}{2\alpha}}\rho\sigma^{\frac{1-\alpha}{2\alpha}})^\alpha\right)&\alpha >1,
\end{cases}
\end{equation}
where for $ \alpha= 1 $, $ \tilde D_1(\rho||\sigma) = \tr (\rho(\log\rho-\log\sigma))$ which follows by demanding continuity in $\alpha$.
To apply these data processing inequalities for CTOs, note that CTOs and TOs are directly related, i.e.
\begin{equation}\label{key}
\rho_S\xrightarrow[\rm CTO]~\rho_S' ~~\iff  \rho_S\otimes \omega_C\xrightarrow[\rm TO]~\rho_S' \otimes \omega_C
\end{equation} 
for some $ \omega_C $. Let $ \mathcal{N}_{\rm TO} $ be the quantum channel such that
$ \mathcal{N}_{\rm TO} (\rho_S\otimes \omega_C) = \rho_S '\otimes \omega_C $.
Then we also know that $  \mathcal{N}_{\rm TO} (\tau_{SC}^\beta) = \tau_{SC}^\beta$, i.e. the thermal state of $ \tau_{SC}^\beta = \tau_S^\beta\otimes\tau_C^\beta$ is of product form (because it is assumed that there is no interaction term in the Hamiltonian of $ \hat H_{SC} $), and is preserved by $ \mathcal{N}_{\rm TO} $. If we know that $ \rho_S\xrightarrow[\rm CTO]~\rho_S' $, then for any quantity $ f $ that satisfies a data processing inequality in Eq.~\eqref{eq:DPI}, there exists $ \omega_C $ such that
\begin{equation}\label{key}
f(\rho_S\otimes\omega_C||\tau_{SC}^\beta) \geq f(\rho_S'\otimes\omega_C||\tau_{SC}^\beta).
\end{equation}
Furthermore, if $ f $ is additive under tensor product, as both variants of the quantum R{\'e}nyi divergences are, we have
$ f(\rho_S||\tau_S)\geq f(\rho_S'||\tau_S) $.

The fact that $ \tilde D_\alpha (\rho_S||\tau_S) $ decreases is insufficient to guarantee that a transition may occur via TO/CTO. For example, instead of using $ \tau_S^\beta $ in the second argument of $ f $, one can consider the quantity
$ \tilde D_\alpha (\rho_S||\mathcal{D}_{\hat H_S}(\rho_S)) $, where $ \mathcal{D}_{\hat H_S} $ is the operation that decoheres $ \rho_S $ in the energy eigenbasis of $ \hat H_S $. Then it is also known \cite{lostaglio2015description} that $ \tilde D_\alpha (\rho_S||\mathcal{D}_{\hat H_S}(\rho_S)) \geq \tilde D_\alpha (\rho_S'||\mathcal{D}_{\hat H_S}(\rho_S')) $ must also be true, and one can find examples where $ \tilde D_\alpha (\rho_S||\tau_S) $ decreases but $ \tilde D_\alpha (\rho_S||\mathcal{D}_{\hat H_S}(\rho_S)) $ does not. Intuitively, this quantifies solely the amount of coherence between distinct energy subspaces in the state $ \rho_S $; and such coherences can only decrease under TOs/CTOs. However, demanding that both quantities are decreasing still does not provide sufficient conditions for a transition.

\subsection{Necessary and sufficient conditions for energy-incoherent states}\vspace{-0.3cm}
If we consider only initial and final states $ \rho_S,\rho_S' $ which are energy-incoherent, then the necessary conditions we saw on $ \tilde D_\alpha (\rho_S||\tau_S^\beta) $ can be extended to become sufficient as well. First of all, note that if $ \rho_S $ and $ \rho_S' $ both commute with $ \tau_S^\beta $, the quantum R{\'e}nyi divergences simplify to their classical decompositions $ D_\alpha (p||q) $, where $ p $ and $ q $ are simply the eigenvalues of respective states. For example, let $\rho_S=\sum_i p_i\ketbra{E_i}{E_i}$, and  $\tau_S^\beta=\sum_i q_i\ketbra{E_i}{E_i}$, where $q_i=e^{-\beta E_i}/Z_\beta$. Then we know that $ \tilde D_\alpha (\rho_S||\tau_S^\beta) =D_\alpha (p||q) $, where
\begin{equation}\label{eq:Dalp}
D_\alpha(p\|q):=\frac{1}{\alpha-1}\ln\left( \sum_i p_i^\alpha q_i^{1-\alpha} \right), ~~\forall \alpha\geq 0.
\end{equation}
Therefore, for two commuting states $ \rho,\sigma $, we will also write $ D_\alpha(\rho||\sigma)=D_\alpha(\eig(\rho)||\eig(\sigma)) $.
In order to specify the state transition conditions, one needs to extend the definition of $ D_\alpha $ for the regime of $ \alpha <0 $, by multiplying Eq.~\eqref{eq:Dalp} with $ \sgn(\alpha) $. For the regime of $ \alpha\geq 0 $, Eq.~\eqref{eq:Dalp} remains unchanged; however for negative $ \alpha $,
\begin{equation}\label{eq:24}
D_\alpha(p\|q):=\frac{1}{1-\alpha}\ln\left( \sum_i p_i^\alpha q_i^{1-\alpha} \right),~~\forall\alpha <0.
\end{equation}
The R{\'e}nyi divergences in Eqs. \eqref{eq:Dalp}, \eqref{eq:24} now collectively determine the possibility of a state transition via CTOs. We state this in terms of the following theorem:
\begin{theorem}\label{thm:2ndlwas}(Second laws for block-diagonal states) Consider a system with Hamiltonian
	$ \hat H_S $, and states $ \rho_S,\rho_S' $ such that $ [\rho_S,\hat H_S]=[\rho_S',\hat H_S]=0 $. Then for any $ \varepsilon >0 $, the following are equivalent: 
	\begin{enumerate}[leftmargin=0.5cm,noitemsep]
		\item $ D_\alpha (\rho_S||\tau_S^\beta)\geq D_\alpha (\rho_S'||\tau_S^\beta) ,~\forall \alpha\in (-\infty,\infty). $
		\item For any $ \varepsilon >0 $, there exists a catalyst $ \omega_C $, and a thermal operation $ \mathcal{N}_{\rm TO} $ such that  $ \mathcal{N}_{\rm TO} (\rho_S\otimes\omega_C) = \rho_S'^\varepsilon\otimes\omega_C $ and $ d(\rho_S',\rho_S'^\varepsilon)\leq \varepsilon$.
	\end{enumerate}
\end{theorem}
\noindent In other words, the transition $ \rho_S\xrightarrow[\rm CTO]~\rho_S' $ can be performed arbitrarily well, if and only if we have $ D_\alpha (\rho_S||\tau_S^\beta)\geq D_\alpha (\rho_S'||\tau_S^\beta) $ as defined in Eqs.~\eqref{eq:Dalp} and \eqref{eq:24}. 
Moreover, if one is willing to use an additional, pure qubit ancilla $ \ketbra{0}{0} $ and return it back $ \varepsilon $-close to its original state, then the requirement on R{\'e}nyi divergences for $ \alpha <0 $ is no longer necessary, because they will be automatically determined by the fulfillment of $ D_0 (\rho_S||\tau_S^\beta)\geq D_0 (\rho_S'||\tau_S^\beta) $. Thererfore, in that case, only the R{\'e}nyi divergences for $ \alpha\geq 0 $ matter.

For an energy-incoherent state $ \rho_S $, one can define \emph{generalized free energies}
\begin{equation}\label{eq:chpt3CTOscc}
F_\alpha (\rho_S,\hat H) := \beta^{-1} \left[ \ln Z_\beta + D_\alpha(\rho_S\|\tau_S^\beta) \right].
\end{equation}
Since these are equivalent to $ D_\alpha $ up to an extra multiplicative and additive constant, we see that $ \rho_S\xrightarrow[\rm CTO]~\rho_S' $ iff $ F_\alpha (\rho_S,\hat H) \geq F_\alpha (\rho_S',\hat H) $. These inequalities are known as the generalized second laws of quantum thermodynamics \cite{secondlaw}. 

While this is a continuous family of second laws, there are three special instances of $ \alpha $ for Eq. \eqref{eq:chpt3CTOscc} that have notable significance. The quantity $F_1$ is defined by demanding continuity in $\alpha$, and is equal to the well-known non-equilibrium free energy in Eq. \eqref{eq:free energy}. Therefore, these generalized second laws contain the standard second law as a special case, but demonstrate that we have more constraints when considering small quantum systems. 

On the other hand, the quantities $ F_0 $ and $ F_\infty :=\displaystyle\lim_{\alpha\rightarrow\infty} F_\alpha $ correspond to the amount of \emph{single-shot distillable work} and \emph{work of formation} \cite{HO-limitations} respectively. In other words, if we adapt the requirements on work as described in Section \ref{subsub:detwork}, then $ F_0 (\rho_S,\hat H_S) $ gives the maximum amount of work extractable $ \Wext $, when considering $ \hat H_W = \Wext \ketbra{1}{1}_W $, such that the transition 
$ \rho_S\otimes \ketbra{0}{0}_W \xrightarrow[\rm CTO]~ \rho_S'\otimes\ketbra{1}{1}_W  $
is possible for some $ \rho_S' $. One can easily show that $ \rho_S' = \tau_S $ obtains the maximum $ \Wext $. Therefore, the maximum amount of single-shot work can always be obtained by thermalizing $ S $ to its surroundings. In fact, for such a transition, a catalyst is not needed; therefore $\Wext$ when maximized over all final states, can already be achieved via TOs. However, if we would require a particular fixed $ \rho_S' \neq \tau_S $ to be achieved, then the corresponding amount of $ \Wext $ is given by optimizing
\begin{equation}\label{key}
\Wext = \inf_{\alpha\geq 0} [F_\alpha(\rho_S,\hat H_S)-F_\alpha(\rho_S',\hat H_S)],
\end{equation}
where the infimum may happen on any $ \alpha $, depending on the specific states $ \rho_S,\rho_S' $. 
Similar to the quantity $ F_0 $, $ F_\infty (\rho_S,\hat H_S) $ gives the minimum amount of work $ W_{\rm form} $ required to create $ \rho_S $ starting from the thermal state $ \tau_S $. More precisely, it gives the minimum possible value of $ W_{\rm form} $ where $ \hat H_W = W_{\rm form} \ketbra{1}{1}_W $, so that the transition $  \tau_S\otimes \ketbra{1}{1}_W \xrightarrow[\rm CTO]~ \rho_S\otimes\ketbra{0}{0}_W $ is possible. Again, one can also achieve this optimal value of $W_{\rm form}$ without a catalyst, since the initial state is thermalized with its environment.

\subsection{Relaxing conditions on CTOs}\vspace{-0.3cm}
Given the continuous, infinite set of generalized second laws that are hard to check, the question naturally arises to whether one can further extend CTOs so that these laws simplify. In particular, one could ask: what happens if we allow for the catalyst to be returned only approximately after the process? This question is also physically motivated, since for realistic scenarios, the catalyst might undergo slight degradation due to uncertainties such as those in
the initial state, or imperfections in the implementation of quantum operations. These factors can induce small, unnoticed changes in a single use of the catalyst, and cause it to gradually lose its catalytic ability.
At the macroscopic scale, the fact that a process is only approximately cyclic has generally
been assumed to be enough to guarantee the second law. However, this is not the case in the microscopic regime, which leads to a phenomenon known as \emph{thermal embezzling}. 
\subsubsection{Thermal embezzling and approximate catalysis}
\vspace{-0.3cm}
At first glance, one might be tempted to demand closeness of $\omega_C'$ to $\omega_C'$ to be quantified in terms of trace distance $d(\omega_C,\omega_C')$, since this quantity tells us how well one can distinguish two quantum states, given the best possible measurement. In terms of the catalyst, one might hence ask 
that $d(\omega_C,\omega_C') \leq \varepsilon$ for some arbitrary small $\varepsilon$. 

However, if there are no additional restrictions on what catalysts are allowed, then the generalized free energies are non-robust against errors induced in the catalyst. This can already be demonstrated in the case where all involved Hamiltonians are fully-degenerate: given a $ d_S $-dimensional system $ S $, for any $ \varepsilon >0 $ error one allows on the catalyst, one can always find a corresponding $ d_C(\varepsilon) $-dimensional catalyst $ \omega_C $, such that any state transition 
$ \rho_S\otimes\omega_C\rightarrow\rho_S'\otimes\omega_C' $
is allowed for all $ \rho_S $ and $ \rho_S' $, while $ d(\omega_C,\omega_C')\leq \varepsilon $. Naturally, the dimension $ d_C $ has to diverge as $ \varepsilon\rightarrow 0 $. Intuitively, one understands this by the fact that maximum entropic difference between $ \omega_C $ and $ \omega_C' $ is not only determined by $ \varepsilon $, but also grows with $ \log d_C $ (as shown in the Fannes-Audenart inequality \cite{audenaert2007sharp}). Therefore, any restriction on trace distance error $ \varepsilon $ can be hidden by choosing a large enough catalyst Hilbert space, so that one extracts a small amount of resource (in terms of energy/purity) that remains relatively unnoticed, and using this resource to enable transitions on the (relatively small) system $ S $.
Thermal embezzling tells us that whenever catalysts are used, it is important to assess changes, however slight (even according to seemingly reasonable measures), on the catalyst. 
There are currently two known ways to further restrict approximate catalysis, in a way that the family of generalized free energies relax to the standard free energy in Eq.~\eqref{eq:free energy} being the only condition for state transitions (for energy-incoherent states):
\begin{enumerate}[leftmargin=0.5cm,noitemsep]
	\item Requiring that the catalyst used must be returned with error
	$	d(\omega_C,\omega_C')\leq {\rm O}\left( \frac{\varepsilon}{\log d_C}\right) $ for any fixed $ \varepsilon >0 $, and $ d_C $ being the dimension of the catalyst \cite{2ndlaw}. This restriction guarantees that although the catalyst is returned with some error, embezzling is avoided since the free energy difference between $ \omega_C $ and $ \omega_C' $ is required to be arbitrarily small.
	\item Allowing correlations to build up in the final reduced state of $ \rho_{SC}' $, such that $ \rho_C' = \omega_C $ is still returned exactly \cite{mueller2017correlating}, while allowing some error $ \varepsilon $ on the final, created state. It is important to note that as the desired $ \varepsilon\rightarrow 0 $ is desired, the required catalyst dimension $ d_C\rightarrow\infty $ as well. Nevertheless, this construction has the appealing effect that the final reduced state $ \rho_C' $ is exactly preserved after enabling the transformation on system $ S $, and therefore $ \rho_C' $ can be used in a second, fresh round of state transitions on $ S' $, as long as $ S' $ is initially uncorrelated from $ S $.
\end{enumerate}
\subsubsection{Approximate system transformations recovering standard free energy in the thermodynamic limit}
\vspace{-0.3cm}
While most literature on TRTs are concerned with exact state transformations, in realistic implementations, we may be satisfied as long as the transition is approximately achieved. 
This has been studied theoretically, for example in the context of probabilistic thermal operations \cite{alhambra2015probability}, and in work extraction protocols when heat/entropy is inevitably produced alongside \cite{aaberg2013truly,Woods2015}. Therefore, it is a natural and physical relaxation of CTOs to consider approximate state transitions: in other words, given initial and target states $ \rho_S $ and $ \rho_S' $, and some error parameter $ \varepsilon $, can we decide if there exists: 
\begin{enumerate}[leftmargin=0.5cm,noitemsep]
	\item a state $ \sigma_S $ such that $ d(\rho_S,\sigma_S) \leq \varepsilon$,
	\item a state $ \sigma_S' $ such that $ d(\rho_S',\sigma_S') \leq \varepsilon$, 
\end{enumerate}
such that $ \sigma_S\xrightarrow[\rm CTO]~ \sigma_S' $?
Another motivation for considering approximate CTOs is to investigate how the macroscopic law of thermodynamics emerges, in the thermodynamic limit. 
We know that if no approximation is allowed, then the generalized second laws do not change at all, since these quantities are additive under tensor product. However, it has been shown that the conditions for approximate CTOs can be given by a family of \emph{smoothed generalized free energies}, denoted as $ F_\alpha^\varepsilon (\rho_S,\hat H)$. Moreover, consider multiple identical systems $ \rho_S^{\otimes n} $ with respect to the joint Hamiltonian $ \hat H_n = \sum_{i=1}^n \hat H_i $. Then, in the limit where $ n\rightarrow\infty $ and $ \varepsilon\rightarrow 0 $, all the smoothed quantities $ F_\alpha^\varepsilon (\rho_S,\hat H_n)$ converge to the standard free energy in Eq.~\eqref{eq:free energy}, namely for all $ \alpha\geq 0 $,
\begin{equation}\label{eq:smoothed converg lim}
\lim_{\varepsilon\rightarrow 0}\lim_{n\rightarrow\infty} \frac{1}{n}  F_\alpha^\varepsilon (\rho_S^{\otimes n},\hat H_n) = F(\rho_S,\hat H).
\end{equation}

\section{Application of second laws: fundamental limits on quantum heat engines}\vspace{-0.3cm}  
\subsubsection{Introduction and motivation}
\vspace{-0.3cm}
One of the initial motivations for the development of thermodynamics was to understand heat engines.
Prior to the development of thermodynamics, engineers had little idea whether there were fundamental limits to the efficiency of heat engines, nor what quantities; if any, such hypothetical limit might depend on. It was Nicolas L{\'e}onard Sadi Carnot who is credited with proposing the first successful theory on maximum efficiency of heat engines in 1824 \cite{carnot}.
In particular, Carnot studied heat engines that extract energy by interacting with different \emph{working fluids}, which are substances (usually gas or liquid) at different temperatures in thermal states; and as such, are defined in terms of their Hamiltonians. He concluded that the maximum efficiency attainable did not depend on the exact nature of the working fluids, but only on their inverse temperatures, $\beta_\ho$,  $\beta_\co$. It was later demonstrated that this maximum efficiency | now known as the Carnot efficiency | is
$ \eta_c=1- \beta_\hot\beta_\cold^{-1} $.
Carnot's results are known to hold in the realm of macroscopic systems | indeed, the form of $ \eta_c $ can be derived from the standard second law of thermodynamics.

Given that the generalized second laws of thermodynamics govern state transitions for microscopic systems, where the laws of large numbers do not apply; 
a natural question is: \textit{does Carnot's famous result still hold?} The importance of this question goes beyond purely academic interests, since people are starting to build atomic scale heat engines \cite{scully2003extracting,nanoscaleHE}. We discuss some recent progress in this direction.

\subsubsection{Setup of a microscopic heat engine}\vspace{-0.3cm}
A heat engine consists of 4 parts; a hot bath, a cold bath, a working body or machine, and a battery where the extracted work can be stored. The Hamiltonians of the systems are $\hat H_\ho$, $\hat H_\co$, $\hat H_\ma$, $\hat H_\bat$, respectively, with a total Hamiltonian $ \hat H_\textup{tot} $ being simply the sum of free Hamiltonians of the individual systems.
A working heat engine should not need any additional systems in order to work. Moreover, allowing for them without explicitly accounting for them, could inadvertently allow for cheating; since one could draw heat or work from them. As such, we only allow for closed dynamics. The machine can be transformed to any state as long as it is returned to its initial state after one cycle. The dynamics during one cycle can be represented by an energy preserving unitary; namely $U(t)$, with $[U(t),\hat H_\textup{tot}]=0$. Here, $t$ is the time per cycle of the heat engine. Note that this formalism allows for an arbitrarily strong interaction term $\hat I_{\co\ho\ma\bat}$ between subsystems, as long as it commutes with $\hat H_\textup{tot}$ to preserve energy. This set-up readily fits into the resource theory framework: the machine acts as a catalyst, while one of the thermal baths is explicitly accounted for in the system, for example the cold bath, $\tau^0_\co$. In the language of CTOs, one cycle of the heat engine is possible, iff $\tau_{\beta_\co}^0\otimes \rho_\bat^0 \xrightarrow[\rm CTO]~\rho_{\co\bat}^1$, where $\rho_{\co\bat}^1$ is the final state of the cold bath and battery. 

As we have seen in Section \ref{Defining Work}, not all measures of energetic changes can act as a suitable quantifier of work. Several different characterizations of work have been proposed, and it is very important to first specify which characterization is considered, when comparing and analysing the maximum efficiency of heat engines.
	For example, it was shown in \cite{P2Nelly} that with a particular characterization called \emph{imperfect} work, one can surpass the Carnot efficiency with this set-up. This is not because microscopic heat engines are superior to macroscopic ones, but because imperfect work detects only the average amount of energy change in the battery, but not the amount of entropy production. On the other hand, at the microscopic scale, one has to be careful not to go to the opposite extreme, since it was proven in \cite{Woods2015} that there exists no final state $\rho_{\co\bat}^1$, where the battery outputs deterministic work, namely $\rho_\bat^0=\ketbra{E_j}{E_j}_\bat,$ $\rho_\bat^1=\ketbra{E_k}{E_k}_\bat,$ $E_k-E_j>0$, which was defined in Section \ref{subsub:detwork}. This motivates the following definition of work, which for all purposes, is as good as deterministic work, but without the technical caveats. Let the final battery state be $\varepsilon$ close to $\ketbra{E_k}{E_k}_\bat$ in trace distance, $d\left( \rho^1_\bat,\ketbra{E_k}{E_k}_\bat\right)=\varepsilon$, with $\varepsilon$ bounded away from one. The work extracted, $W_\ext= E_k-E_j$, is \emph{near perfect} if there exists an infinite sequence of heat engines such that, for all $p>0$, 
\be\label{eq:nearperfectwork}
0<\frac{\Delta S}{W_\textup{ext}}\leq p
\ee
holds for some proper subset of the sequence, where $\Delta S$ is the increase in von Neumann entropy of the battery.

\subsubsection{Heat engine efficiency}\vspace{-0.3cm}
The efficiency of a heat engine is defined as 
$ \eta:=\frac{W_{\rm ext}}{\Delta H} $,
where $\Delta H$ is the amount of heat drawn from the hot bath, namely $\Delta H= \tr(\hat H_\ho \rho_\ho^0)-\tr(\hat H_\ho \rho_\ho^1)$. Since the machine's final and initial states are the same after one cycle, and the initial state of the cold bath is fixed; due to total mean energy conservation, $\Delta H$ is a function of $\rho_\cold^1$ and $W_\textup{ext}$. We can then define the maximum achievable efficiency in the nanoregime $\eta^\textup{nano}$ as a function of the final state of the cold bath $\rho_\cold^1$. More precisely,
\begin{align}
	&\eta^\textup{nano}
	(\rho_\cold^1)\label{eq:max nano as function fo cold bath}
	:=\sup_{W_\textup{ext}>0} \eta(\rho_\cold^1) \quad \textup{ subject to }&\\
	&F_\alpha(\rho^0_\battery\otimes\tau_\cold^0
	)\geq F_\alpha(\rho_{\stcold\bat}^1
	)\quad \forall \alpha\geq 0,&\label{eq:second laws line}
\end{align}
where $F_\alpha$ are defined in Eq. \eqref{eq:chpt3CTOscc}.

We are interested in answering the question: under what conditions, if any, can Carnot efficiency be achieved? It is known that in regimes where we only have the usual 2nd law, Carnot efficiency can only be achieved when the free energy remains constant during the transition. This is the special case in Eq. \eqref{eq:second laws line} corresponding to
\be \label{eq:F1 eq}
F_1(\rho^0_\battery\otimes\tau_\cold^0
)=F_1(\rho_{\stcold\bat}^1
).
\ee
Due to sub-additivity of entropy, $F_1(\rho_{\stcold\bat}^1)>F_1(\rho_{\co}^1\otimes\rho_{\bat}^1)$ iff $\rho_{\stcold\bat}^1$ is correlated, so correlations in the final state cannot help one to achieve the Carnot efficiency if one assumes finite-dimensional catalysts. In addition, it can be shown that for near perfect work, that for constant $\Delta C$, the state of the cold bath which maximises the efficiency $\eta$ when only $F_1$ is required to be non-increasing, is when $\rho_{\co}^1$ is a Gibbs state. Furthermore,  when $\rho_{\co}^1$ is a Gibbs state, equality in Eq. \eqref{eq:F1 eq} can only be achieved in the limiting case in which the temperature of the final state of the cold bath approaches the initial cold bath temperature from above. We call this limit in which the Carnot efficiency can be reached, according to the standard second law, the \textit{quasi-static heat engine} limit. The natural question, is whether or not a quasi-static heat engine can achieve the efficiency in the microscopic regime. For this, not only will $F_1$ have to be non-increasing, but rather all generalised free energies in Eq. \eqref{eq:second laws line} will have to be.

To answer this question, we specialise to the case in which the cold bath Hamiltonian is comprised of $n$ qubits, with the $i$-th qubit energy gap denoted $\Delta E_i$; and introduce the parameter $\Omega$
\be 
\Omega:=\min_{i=1,2,\ldots,n} \left(\beta_\co-\beta_\ho\right)\frac{\Delta E_i}{Z_i\, \me^{\beta_\co\Delta E_i}}
\ee
where $Z_i$ is the partition function of the $i$-th qubit.

\begin{theorem}(From \cite{Woods2015})\label{thm:quasi static eff}
	A quasi-static heat engine extracting near perfect work can achieve the Carnot efficiency, iff $\Omega \leq 1$, otherwise only a strictly smaller efficiency is achievable, specifically
	\be \label{eq:eta nano}
	\eta^\textup{nano}=\left(1+\frac{\beta_\hot}{\beta_\cold-\beta_\hot}\Omega\right)^{-1}.
	\ee 
\end{theorem}
To see that Eq.~\eqref{eq:eta nano} is indeed less that the Carnot efficiency when $\Omega>1$, observe that $\eta^\textup{nano}$ achieves the Carnot efficiency for $\Omega=1$ and that it is monotonically decreasing in $\Omega$.

Theorem \ref{thm:quasi static eff} has significant consequences for Carnot's famous results about the universality of the maximum efficiency of heat engines. Namely, that it does not apply at the microscopic scale when at least one of the baths is finite, since not all Hamiltonians of the thermal baths | the working fluids, in the language of Carnot | 
can achieve Carnot efficiency. This said, observe how for any pair $\beta_\ho$, $\beta_\co$; when energy gaps of the cold bath are sufficiently small such that $\Omega<1$, then Carnot efficiency is again achievable; even for this finite heat bath.

\section{Developments and open questions in thermodynamic resource theories}\vspace{-0.3cm}
The simplicity of a resource theoretic approach towards quantum thermodynamics is its main appeal and power, although it may at the same time be its weakness. 
Given the large amount of past endeavours in modeling thermodynamical interactions, via Master/Linblad equations \cite{gemmer:book, kosloff2013quantum,Alicki1979,ent_enchance_cooling,PhysRevLett.105.130401,PhysRevE.87.012120,PhysRevX5031044} (especially those involving strong coupling system-bath Hamiltonians \cite{PhysRevA.79.010101,peter2008finite,ingold2009specific}), time dependence in Hamiltonians (such as quenches \cite{PhysRevLett.96.136801,cramer2008exact}) etc, the question arises as to how TRTs relate to these approaches. 
Since the primitive version of noisy/thermal operations, various works have endeavoured to extend and connect the usage of TRTs with other approaches. In particular, we list
some important results to date:\vspace{-0.3cm}
	\subsubsection{Inclusion of catalysts}\vspace{-0.3cm}
	Thanks to the understanding of CTOs, any experimental apparatus used to implement TOs can now be modelled as additional quantum systems, and be included into the framework of thermal operations. 
	The criteria and guidelines for choosing appropriate catalyst states have been studied, including inexact catalysis, catalysts that may correlate with the system \cite{mueller2017correlating}, and energy-coherent catalysts \cite{aaberg2014catalytic}. 
	Despite increasing knowledge on this subject, we still lack a clear consensus on the full role of catalysts. In light of thermal embezzling, a question arises as to how one may quantify/justify errors on the catalyst. Moreover, for current results on the subject, the state of a useful catalyst still depends strongly on the desired transformation (initial and final states of the system), and it would be desirable to have generic catalysts which are not only useful for particular transformations, but at least for a subset of processes.\vspace{-0.3cm}
\subsubsection{Implementation of TOs via autonomous models}\vspace{-0.3cm}
	A fully autonomous model of thermodynamical interactions would simply be described by a time independent Hamiltonian. Such a scenario usually depicts a naturally arising physical process, since systems simply evolve spontaneously according to a fixed Hamiltonian without any control from an observer. Since unitary operations such as those in TRTs are applied at a specific time, one can understand them to be implemented by using an additional quantum system as a clock, and a time independent Hamiltonian over the system and clock Hilbert-spaces, controlled on the clock's state. 
	One of the initial models to attempt to address this issue \cite{malabarba2015clock}, suffers from the flaw of having a clock with no ground state (thus infinite energy), where no physical realization is possible. This has the unphysical consequence of never suffering any back-reaction of the clock due to the implementation of unitaries regardless of how quickly there are implemented. 
	Another approach \cite{woods2016autonomous}, while not suffering from these issues; has so-far proven inconclusive about the true thermodynamical cost, and whether there is extra cost unaccounted for in controlling quantum systems by applying autonomously energy preserving unitaries. Despite exciting progress that has been made, this is still an important ongoing area of research.\vspace{-0.3cm}
\subsubsection{Equivalence to other operations} \vspace{-0.3cm}
	In order to study the fundamental limits to thermodynamics, TOs consider a large set of operations: the experimenter is allowed to use an arbitrary thermal bath, and any energy-preserving unitary. This allows for the derivation of statements holding for full generality, however, one may question if there exists simple operations that would also achieve the full power of TOs.
	Recently it has been shown that all TOs for energy-incoherent states can be accomplished via a finite number of \emph{coarse operations} \cite{perry2015sufficient}, where the latter involves qubit baths, level transformations and thermalization. This partially reconciles the TRT framework with various other approaches in quantum thermodynamics \cite{egloffdahlsten,del2011thermodynamic,aaberg2013truly,Crooks1998,infoerasure}. Moreover, recent studies have further identified subsets of TOs that have clear experimental realizations \cite{lostaglio2016elementary}. These efforts pave the way to designing protocols that achieve the optimal state transitions as predicted by TOs. See \cite{YungerHalpern2017} for a more detailed review of recent attempts to realise resource theoretic approaches to quantum thermodynamics. 
	\subsubsection{Relation with fluctuation theorems}\vspace{-0.3cm}
	A distinct approach to quantum thermodynamics known as fluctuation relations has been independently progressing in parallel to the development of TRTs \cite{secondlaweqnineq2011,nonequifluccount2009,revmodphyscolloquium2011QFR,stochasticthermo2012}. A handful of experimental verifications for these relations have been demonstrated, both in the classical \cite{denis2002} and quantum regime \cite{utsumi2010,exp_reconst_QFR}. However, the stark conceptual differences between fluctuation relations and TRTs have prevented them, so far, from being connected. Recently, the possibility of connecting both to form a harmonious picture of thermodynamics has been explored \cite{epsdetwork,alhambra2016second}. Should this be achieved, it would provide us potential means to experimentally investigate the predictions of TRTs by making use of fluctuation relations demonstrations.

\acknowledgements
N.N. acknowledges funding from the Alexander von Humboldt foundation.
M.W. would like to acknowledge the COST MP1209 network ``Thermodynamics in the quantum regime", to which he was a member and from which part of the research reviewed in this chapter was made possible.


%

\end{document}